# Spin-orbit torque in completely compensated synthetic antiferromagnet

P. X. Zhang,[1,*] L. Y. Liao,[1] G. Y. Shi,[1] R. Q. Zhang,[1] H. Q. Wu,[2] Y. Y. Wang,[3] F. Pan,[1] and C. Song[1,†]

[1]Key Laboratory of Advanced Materials (MOE), School of Materials Science and Engineering, Tsinghua University, Beijing 100084, China.

[2]Institute of Microelectronics, Tsinghua University, Beijing 100084, China.

[3]Department of Physics, Beihang University, Beijing 100191, China

Synthetic antiferromagnets (SAF) have been proposed to replace ferromagnets in magnetic memory devices to reduce the stray field, increase the storage density and improve the thermal stability. Here we investigate the spin-orbit torque in a perpendicularly magnetized Pt/[Co/Pd]/Ru/[Co/Pd] SAF structure, which exhibits completely compensated magnetization and an exchange coupling field up to 2100 Oe. The magnetizations of two Co/Pd layers can be switched between two antiparallel states simultaneously by spin-orbit torque. The magnetization switching can be read out due to much stronger spin-orbit coupling at bottom Pt/[Co/Pd] interface compared to its upper counterpart without Pt. Both experimental and theoretical analyses unravel that the torque efficiency of antiferromagnetic coupled stacks is significantly higher than the ferromagnetic counterpart, making the critical switching current of SAF comparable to the conventional single ferromagnet. Besides adding an important dimension to spin-orbit torque, the efficient switching of completely compensated SAF might advance magnetic memory devices with high density, high speed and low power consumption.

---

[*] Present address: Department of Electrical Engineering and Computer Science, Massachusetts Institute of Technology, Cambridge, Massachusetts 02139, USA
[†] E-mail: songcheng@mail.tsinghua.edu.cn

## I. INTRODUCTION

The magnetic tunnel junction (MTJ) stands out as a seminal spintronic material due to its high magnetoresistance and extensive applications in electronic devices, such as magnetic random access memory (MRAM) and high sensitivity sensor [1–4]. Different from the reference layer composed of synthetic antiferromagnets (SAF), the free layer of MTJ is generally a single ferromagnetic layer to ensure effective switching by spin-transfer torque. However, a pressing demand for higher storage density and smaller junction size depends on further reducing stray field and enhancing thermal stability. Antiferromagnets (AFM) with zero net magnetic moment, strong anti-interference performance and ultra-fast switching speed have a potential competitiveness in stable and faster information storage. These advantages enable AFM to develop from a traditional supporting layer in an exchange bias system to a functional material in antiferromagnet spintronics [5–10]. Nevertheless, signal writing and reading in antiferromagnetic storage layers remained difficult, except the recent observation of current-driven Néel-order spin-orbit (field-like) torque switching in CuMnAs [11]. However, such a torque switching is limited in the AFMs with a specific spin-sublattice (the known materials are only CuMnAs and $Mn_2Au$) [11–14], while the antidumping torque-induced switching was very recently observed in biaxial antiferromagnets [15]. But the compatibility of these AFMs with device integration still needs further demonstration.

SAF formed by exchange-coupled ferromagnetic bilayers combines the advantages of zero stray field and high stability from AFMs, as well as easy reading and writing characteristics from ferromagnets, making it a promising candidate for information storage [16,17]. Utilizing SAF as free layer in MTJ has been proposed to enhance the thermal stability and reduce switching critical current [18,19]. Recently, an emergent method for manipulating magnetization, namely spin-orbit torque (SOT), was demonstrated to switch magnetization and drive domain wall motion with higher efficiency and lower power consumption than spin transfer torque [20–28]. Despite that SOT induced SAF switching has been extensively studied in Co/Ru/[Co/Pt] and CoFeB/Ta/CoFeB systems [29,30], determination of SOT efficiencies in SAF still remains elusive. Meanwhile, it is significant to enhance the exchange coupling field and reduce the uncompensated spins for an "ideal" SAF. However, these features intrinsically go against the magnetization switching and signal readout, where we confront a dilemma. Therefore we must make sure that the spin torque efficiency is enhanced in the SAF, and the magnetization switching signal could be read out by an interfacial design.

The experiments below demonstrate the SOT in Pt/[Co/Pd]/Ru/[Co/Pd] SAF structure with completely compensated magnetization and wide antiferromagnetic coupling plateau, where the critical switching current is comparable to the ferromagnetic coupled one due to the greatly enhanced torque efficiency. Because of much stronger spin-orbit coupling at bottom Pt/[Co/Pd] interface compared to its upper counterpart without Pt, the magnetization switching can be read out.

## II. EXPERIMENTAL

Ta(20)/Pt(40)/Co(4)/Pd(5)/Co(4)/Ru($t_{Ru}$)/Co(4)/Pd(5)/Co(4)/Pd(20) ($t_{Ru}$ = 2, 3, 4, 5, 6, 7, 8 and 9, units in Ångström) stacks were deposited on thermal oxidized Si/SiO$_2$ substrate via e-beam evaporation at a base pressure of $5 \times 10^{-9}$ Torr. The sample layout is displayed in Fig. 1(a). In the stack structure, the spin current generated by the spin Hall effect (SHE) in the heavy metal Pt would flow upward and switch the magnetization of the bottom [Co/Pd] layer. The deposition rates for Ta, Pt, Ru and top Pd layer films were ~0.1 Å/s and the deposition rates for Co and interlayer Pd were kept ~0.05 Å/s for more precise control of the film thicknesses. Then the multilayers were patterned into Hall bar devices with channel width of 3 μm utilizing photolithography and Ar ion etching. After that, the Ti(10)/Au(100) (units in nanometers) electrodes were prepared by e-beam evaporation and lift-off process. The magnetization measurements were carried by superconducting quantum interference device (SQUID). The anomalous Hall effect (AHE) and current-induced magnetization switching were carried out by four point measurements in a Hall cross with channel width of 3 μm at room temperature. The measurement configuration is sketched in Fig. 1(b). By rotating samples, the external magnetic field could be applied in any direction in yz plane.

## 3. RESULTS AND DISCUSSION

We first show in Fig. 2 the out-of-plane hysteresis loops of the stack films with Ru spacer layer with different thickness. For $t_{Ru}$ = 3 Å [Fig. 2(a)], the magnetization curve exhibits a square loop, indicating the perpendicular magnetic anisotropy (PMA) of the stack films [31]. No separate switching field for the bottom and upper [Co/Pd] layers is observed, i.e., they switch together, reflecting the ferromagnetic coupling between them. The situation turns out to be dramatically different for $t_{Ru}$ = 6 Å. Separate switching fields for the two [Co/Pd] layers are observed in Fig. 2(b). For example, in the descend branch, switching fields are 880 Oe and −1050 Oe for the soft upper and hard bottom [Co/Pd] [31], respectively. Regardless of exchange coupling, the lower Co/Pd layer is always harder than the upper Co/Pd layer, in the sense that the upper layer always switches first. This can be explained by the larger PMA of the bottom layer [31]. The difference in perpendicular magnetic anisotropy is the result of the particular sample growth condition. PMA of our [Co/Pd] grown on FCC Platinum layer is usually stronger due to strain and interfacial effect, compared to [Co/Pd] grown on HCP Ruthenium layer. These two [Co/Pd] layers are antiferromagnetically coupled with the exchange coupling field ($H_{ex}$) of ~970 Oe. Remarkably, the antiferromagnetic coupling is further enhanced when the Ru film thickness increases to 7 Å [Fig. 2(c)]. A wide antiferromagnetic coupling plateau is ~2100 Oe, while the magnetic moments of the two perpendicular magnetized [Co/Pd] layers separated by the Ru spacer are completely compensated, resulting in the nearly zero-moment at the whole plateau. This antiferromagnetic coupling is greatly reduced for $t_{Ru}$ = 8 Å with $H_{ex}$ ~ 800 Oe in Fig. 2(d), followed by the sample of $t_{Ru}$ = 9 Å in Fig. 2(e) with negligible antiferromagnetic coupling. $H_{ex}$ is summarized in Fig. 2(f), where the peak of antiferromagnetic coupling occurs at $t_{Ru}$ = 7 Å with $H_{ex}$ up to 2100 Oe. This variation is consistent to the previous observation of periodic feature of interlayer coupling induced by a Ruderman-Kittel-Kasuya-Yosida (RKKY) exchange interaction [16]. The observation of high antiferromagnetic coupling field indicates the present sample is immune to sizable magnetic perturbation, ensuring the data stability, and its completely compensated moments, which lead to negligible stray field, is beneficial to enhance storage density.

We then address the question whether it is possible to read and write signals in the samples with completely compensated moments. Figure 3(a) presents the AHE curves of the typical [Co/Pd]/Ru/[Co/Pd] samples with $t_{Ru}$ = 6, 7, 8, and 9 Å. The experimental configuration is shown in Fig. 1(b) with the external magnetic field ($H$) applying perpendicular to the film plane and a DC current of 500 μA. The most striking feature for this figure is that all the samples show AHE curves with the wide hysteresis window associated with the antiferromagnetic coupling plateau in Fig. 2. In this scenario, two overlapping states "↑↓" and "↓↑" at the antiferromagnetic coupling plateau of the magnetization curves are successfully separated in electrical measurements, which is significant for the application in memories and sensors with a combination of data stability and writable/readable capabilities. In general, AHE is proportional to the $z$-axis component of magnetization, thus it seems that this rule is in conflict with the observation of clear AHE signals. However, this is not the case for the SAF. The anomalous Hall resistivity is expressed by $\rho_{AH} = 4\pi(R_S^b M_z^b + R_S^u M_z^u)$ [32], where $R_S^b$ and $R_S^u$ are the anomalous Hall coefficient of the bottom and upper layer, respectively, while $M_z^b$ and $M_z^u$ are the corresponding z-axis components of magnetization. The anomalous Hall coefficient follows $R_S^b > R_S^u$ due to the stronger spin-orbit coupling at bottom Pt/[Co/Pd] interface compared to its upper counterpart without Pt [33], which is different from $\Delta M_z = M_z^b + M_z^u$ with opposite $M_z^b$ and $M_z^u$ [31]. In addition, there might also be some minor contribution to the asymmetric AHE from uneven current distribution.

The detectable AHE curves serve as a basis for the SOT switching measurement. For this experiment, Hall resistance ($R_H$) was recorded during the scanning of a current ($I$) applied to the y-axis of the Hall cross with an external magnetic field along +y direction ($\beta = 0°$), and the $R_H$–$I$ data are presented in Fig. 3(b). There are three striking features for these SOT switching curves. First, a gradual switching feature induced by current is observed for all the SOT curves [31], which is quite characteristic for multi-domain switching [34–37]. Second, the two [Co/Pd] layers with antiferromagnetic coupling can be switched between two anti-parallel states simultaneously through the SOT. Only Pt has significant contribution to the SOT because of its strong spin-orbit coupling. Considering that the spin diffusion length of Co, Pd and Ru are only approximately 1.2 nm [38], 2 nm [39], and 4 nm [40] respectively, the upper [Co/Pd] layer would be switched mostly due to the exchange coupling. Note that the high/low resistance state of the $R_H$–$I$ curve around zero-current is lower than its AHE counterpart in Fig. 3(a), which could be ascribed to the canting of out-of-plane magnetic moments by the large assist magnetic field [31]. Third, the critical switching current density ($J_C$) of $t_{Ru} = 7$ Å with strong antiferromagnetic coupling is less than two times of the $J_C$ of $t_{Ru} = 9$ Å sample, though the exchange coupling field $H_{ex}$ of the former is more than 30 times of the later.

If the current density corresponding to 50% switching is set for $J_C$ and the current is assumed to be uniformly distributed in 3 μm × 11 nm cross-section, $J_C$ fluctuates just in a small scale for this series of samples, that is, samples exhibit $J_C$ = 1.5~2.5 × 10$^7$ A/cm$^2$. This finding reveals that SOT is a comparatively effective method to write signal in SAF, which is immune to the external magnetic field. An inspection of the experiments reveals that the current-driven SOT switching only saturates when the assist in-plane magnetic field is comparatively large, i.e., 4 kOe for the $t_{Ru}$ = 7 Å sample and 3 kOe for the other samples in Fig. 3(b). The $t_{Ru}$ = 6–8 Å samples could not show a full and deterministic SOT switching when the assisting field is lower than 1 kOe [31]. Obviously, this behavior is different from the conventional ferromagnetic systems [20–22,24–27], whereas it is similar to the case of ferrimagnets, such as Ta/Co$_{1-x}$Tb$_x$/Ru and Pt/Co$_{1-x}$Gd$_x$/TaO$_x$ [41,42]. Nevertheless, the present SAF shows the relative advantage on stability and compatibility, compared to the ferrimagnets with strong dependence of compensated point to temperature. Here we should note that SOT switching in micrometer-size sample by DC current are often aided by heating. Heating reduces the energy barrier and lowers the threshold current. Although our switching is inevitably heat-assisted, the heating effect on exchange coupling strength should be small within our current range [29]. Furthermore, the SOT efficiency measurement below does not require large current, and it demonstrates that even not assisted by heating, SAF is still possible to be switched by a comparable current as conventional ferromagnet.

When an in-plane assist field exists, the SOT switching efficiency of SAF tells us that how much out-of-plane magnetic field is equivalent to the spin torque in terms of magnetization switching [24]. It can be quantitatively characterized by the effective magnetic field ($H_{eff}$). We record the Hall resistance when a fixed magnitude external field is rotating in yz plane. Figure 4(a) shows representative $R_H$–$\beta$ curves of the $t_{Ru}$ = 6 Å sample around $\beta$ = 0° and 180°. These curves were measured at $H_{ext}$ = 3 kOe and current $I$ = ±3 mA. The most striking feature is the obvious opposite horizontal shift for the positive and negative current. This is because the magnetization switching is motivated by a combination of SOT effect and $z$-component of the external field. For details, when the current of $I = 3$ mA is applied, the spin current from the Pt layer is injected into the bottom [Co/Pd] layer, which acts as an effective field on the $+z$ direction to contribute a part of $H_z$, giving rise to the left shift, while the current of $I = -3$ mA does the opposite. As a result, the shift of the angle $\Delta\beta$ as a function of applied current $I$ at both $H = \pm 3$ kOe is presented in Fig. 4(b). Note that $\Delta\beta$ varies linearly with $I$, and changes its sign with the direction of $H_{ext}$, which coincides with the characteristics of the SOT induced effective field [43,44]. Meanwhile, the average value of $\Delta\beta$ obtained by both the positive and negative $H$ experiments would enhance its accuracy, which is used for the calculation of the effective field $H_{eff}$ as below.

The situation differs dramatically when $t_{Ru}$ increases to 9 Å with negligible antiferromagnetic coupling. The experiments were carried out with the identical procedure as the $t_{Ru} = 6$ Å sample. Remarkably, the shift of the angle $\beta$ for the $I = \pm 3$ mA curve becomes much smaller [Fig. 4(c)], revealing that the effective field is reduced in this sample. As a consequence, the efficiency of effective field $\chi = H_{eff}/J$ as a function of in-plane external field for three typical samples of $t_{Ru} = 6$, 7 and 9 Å and a control sample, which consists of only a [Co/Pd] simple ferromagnet (SF), i.e. Ta(20)/Pt(40)/Co(4)/Pd(5)/Co(4)/Ru(20) (units in Ångström), are summarized in Fig. 4(d). The current density $J$ is calculated assuming a uniform current distribution. $H_{eff}$ is obtained in the following procedure: when the rotated angle $\beta$ is small, the $z$-component $H_z = H_{ext} \sin\beta \approx \beta H_{ext}$, and then $H_{eff}$ is calculated through $H_{eff} = H_{ext}\Delta\beta$. $H_{eff}$ of the samples with strong antiferromagnetic coupling are greatly enhanced, i.e., for $H = 3$ kOe, the corresponding $\chi$ are about 11, 22, 3 and 4 Oe/($10^6$A/cm$^2$) for $t_{Ru} = 6$, 7, and 9 Å samples as well as a control sample with only a [Co/Pd] SF, respectively, indicating that the antiferromagnetic coupling would affect $\chi$ by a factor of six in the SAF samples. Note that deterministic SOT switching just occurs in a certain range of assisting external field: too small $H_{ext}$ cannot assist effective switching, while too high $H_{ext}$ drives the moments to cant toward in-plane. An inspection of the curves shows that $\chi$ almost get saturated when $H_{ext}$ is up to 1 kOe for the $t_{Ru}$ = 9 Å sample, in contrast to almost 3 kOe for strong coupled sample with $t_{Ru}$ = 6 Å, while the $t_{Ru}$ = 7 Å sample with the strongest antiferromagnetic coupling exhibits the maximum $\chi$ of ~22 Oe/($10^6$A/cm$^2$) at 4 kOe [31].

To reaffirm whether the method used above for measuring $H_{eff}$ is reliable, the spin Hall angle is then calculated based on the obtained $H_{eff}$. According to the multi-domain SOT switching mechanism, for conventional ferromagnetic structure, SOT efficiency $\chi = H_{eff} / J_C$ (where $J_C$ represents the current density in the heavy metal layer) is given by [41],

$$\chi = \frac{\pi}{2} \frac{\hbar \xi_{DL}}{2e\mu_0 M_S t_{FM}} \cos\Phi \tag{1}$$

where $M_S$ is the saturated magnetization, $t_{FM}$ parameters the ferromagnetic layer thickness, $\xi_{DL}$ is the effective spin Hall angle, $\Phi$ is the angle between the central moment of the domain wall (later referred as "domain wall moment") and the current, and $\cos\Phi$ is equal to 1 as the SOT efficiency is saturated, as well as $\hbar$, $e$, and $\mu_0$ are Planck constant, elementary charge, and permeability of vacuum, respectively. This equation is applicable to not only conventional ferromagnet sample, but also our $t_{Ru}$ = 9 Å sample with negligible AF coupling, where the bottom layer switches nearly free of the upper layer. Use these parameters for $t_{Ru}$ = 9 Å sample: $M_S$ = 1500 emu/cm$^3$, $t_{FM}$ = 0.4 nm, and $\chi = 10 \frac{Oe}{mA} = 3.45 \frac{Oe}{10^6 A/cm^2}$. The $M_S$ is slightly higher than the bulk value because of thickness uncertainty. $\xi_{DL}$ calculated by Eq. (1) is 0.08. Only damping-like torque has contribution to this equation [24]. Considering that the spin diffusion length of Pt is $\lambda_{SF}$=1.4 nm [22], the spin Hall angle in our case is given by

$$\theta_{\mathrm{SH}} = \frac{\xi_{\mathrm{DL}}}{1-\mathrm{sech}(t_{\mathrm{Pt}}/\lambda_{\mathrm{SF}})} = 0.09$$, assuming a transparent Pt/Co interface. This value is within the extensively accepted range and demonstrates the effectiveness of our method. This value could be considered as the intrinsic spin Hall angle for the SAF system, which does not vary with antiferromagnetic coupling strength. Nevertheless, the effective spin Hall angle calculated by Eq. (1) with the bottom layer saturated magnetization $M$s, thickness $t_{\mathrm{FM}}$ and saturated spin torque efficiency χ does vary with antiferromagnetic coupling strength: the effective spin Hall angle $\xi_{\mathrm{eff}}$ equals to 0.22, 0.47, and 0.08 for $t_{\mathrm{Ru}}$ = 6, 7, and 9 Å samples, samples, respectively.

The behavior of spin-torque efficiency under different applied field can be explained as follows. Because of the strong spin-orbit coupling and the resultant Dzyaloshinskii-Moriya interaction (DMI) at the bottom Pt/Co interface, as shown in Fig. 5(a), the bottom layer domain wall "↑←↓" and "↓→↑" are Néel-type with left hand chirality [24], leading to parallel movements under the spin-torque. We label domain walls by the bottom layer because its chirality is more important. Such chirality would be imprinted to the upper layer ascribed to the interlayer antiferromagnetic coupling. Under large external field, the moments of both "↑→↓" and "↓→↑" domain walls are realigned in +$y$ direction [Fig. 5(b)], leading to opposite movements under the spin-torque.

Based on the SOT switching equivalent force proposed by Ref. [43] and assuming coupled domain wall motion, we get the z-direction effective field of spin-orbit torque in SAF structure approximately as [31],

$$\chi = \frac{\pi}{2}\frac{\hbar\xi_{\mathrm{DL}}}{2e\mu_0 t_{\mathrm{b}}\Delta M_z}\cos\Phi \qquad (2)$$

Where $t_{\mathrm{b}}$ is the bottom layer thickness, and $\Delta M_z = M_z^{\mathrm{b}} + M_z^{\mathrm{u}}$ is the net z component magnetization of bottom and upper layer, respectively, which is determined by macrospin model, as shown in Fig. 5(c). Due to the larger out-of-plane anisotropy of the bottom layer, $\Delta M_z$ has the same sign as $M_z^b$. This equation quantify the driving effect on SAF domain wall from spin torque. The reason for large $\chi$ is not due to increasing effective spin Hall angle $\xi_{DL}$ which actually does not vary, but due to the small $\Delta M_z$. This is very similar to the compensated ferrimagnet case where $M_S$ becomes small [41]. For the whole system, both up-to-down domain walls and down-to-up domain walls (labeled with the moments in the bottom layer) should be considered, i.e., we should replace $\cos\Phi$ with $(\cos\Phi+\cos\Phi')/2$, where $\Phi$ and $\Phi'$ are the angles between the up-to-down and down-to-up domain wall moments and the current as shown in Fig. 5(d).

To derive the $H_{ext}$ dependency of $\chi$, we start with the calculation of $\cos\Phi$, because in general the domain wall is a mixture of Néel and Bloch types due to the competition between various energy terms. In the collective domain wall model, for the up-to-down domain wall, the total domain wall energy [44,45] is expressed as,

$$\sigma_{DW}(H_{ext},\Phi,\psi) = \sigma_b + \sigma_u + 2K_D\lambda(\cos^2\psi + \cos^2\Phi) - \pi\lambda M_b(H_{ext} + H_{DMI}^b)\cos\Phi \\ -\pi\lambda M_u(H_{ext} - H_{DMI}^u)\cos\psi + \pi\lambda J_{ex}\cos(\Phi-\psi) \quad (3\text{-a})$$

where $\sigma_b$ and $\sigma_u$ are the Bloch-type domain wall energy densities of the bottom and upper domain wall, respectively, $K_D$ is the domain wall anisotropy energy density, $\lambda$ is the domain wall width, $\Phi$ and $\psi$ are the angles between the bottom and upper domain wall moments and the current, respectively, $H_{DMI}^b$ and $H_{DMI}^u$ are the DMI effective fields of the bottom and the upper layers, respectively. A similar equation can be written for the down-to-up domain wall as

$$\sigma_{DW}(H_{ext},\Phi',\Psi') = \sigma_b + \sigma_u + 2K_D\lambda(\cos^2\psi' + \cos^2\Phi') - \pi\lambda M_b(H_{ext} - H_{DMI}^b)\cos\Phi' \\ -\pi\lambda M_u(H_{ext} + H_{DMI}^u)\cos\psi' + \pi\lambda J_{ex}\cos(\Phi'-\psi') \quad (3\text{-b})$$

To show the origin of the high required assist-field to achieve the deterministic SOT switching, we simplify Eq. (3) with the assumption that domain wall moments can only choose $+y$ or $-y$ direction, i.e., $\Phi$, $\psi$, $\Phi'$ and $\psi'$ can either be 0 or $\pi$. In the studied SAF structure the $H_{DMI}^{u}$ is provided by Pd/Co interface and $H_{DMI}^{b}$ is provided by Pt/Co interface, therefore $H_{DMI}^{u} < H_{DMI}^{b}$. Thus, Eq. (3) can be solved as

$$\Phi = \psi' = \pi, \psi = \Phi' = 0, 0 < H_{ext} < H_{DMI}^{u} + H_{ex} \quad (i)$$
$$\Phi = \pi, \psi' = \psi = \Phi' = 0, H_{DMI}^{u} + H_{ex} < H_{ext} < H_{DMI}^{b} + H_{ex} \quad (ii) \qquad (4)$$
$$\Phi = \psi' = \psi = \Phi' = 0, H_{ext} > H_{DMI}^{b} + H_{ex} \quad (iii)$$

The critical point between condition (i) and (ii) is $H_{ext} = H_{DMI}^{u} + H_{ex}$, when DW2 in Fig. 5(a) switches from $\pi$ to 0. And the critical point between condition (ii) and (iii) is $H_{ext} = H_{DMI}^{b} + H_{ex}$, when DW1 in Fig. 5(a) switches from $\pi$ to 0. In both condition (i) and (ii) we have $\Phi = \pi$ and $\Phi' = 0$, so the moments of up-to-down ($\uparrow \leftarrow \downarrow$) and down-to-up ($\downarrow \rightarrow \uparrow$) domain walls are opposite, therefore the domain walls propagate in parallel and no switching occurs, characterized by a zero SOT efficiency. In condition (iii), however, both the central moments of up-to-down ($\uparrow \rightarrow \downarrow$) and down-to-up ($\downarrow \rightarrow \uparrow$) domain walls point to $+y$ direction, leading to the opposite movements and the domain expansion. Therefore, the characteristic assist-field for SAF deterministic SOT switching is $H_{DMI}^{b} + H_{ex}$, which is relatively larger compared to conventional ferromagnet with characteristic assist-field $H_{DMI}^{b}$. The presence of these critical points is verified by our simplified Néel wall micromagnetic model [31].

In reality, the switching between condition (i), (ii) and (iii) are not sudden changes from $\pi$ to 0, but gradual rotations under the different competing energies. To use the full form of Eq. (3) to describe this process, we obtain $H_{DMI}$ and $K_D$ by fitting the spin-torque efficiency data of simple ferromagnet in Fig. 4(d) [31]. Corresponding data are presented in Fig. 5(e). The results are in consistent with the measurement results shown in Fig. 4(d), except a dip in small field region in SAF case, caused by the counter rotation of Φ' with the π to 0 rotation of ψ' under small external field. This behavior does not appear in our experiment, possibly because some extrinsic factors such as defect pinning govern the behavior in small field region.

A significant difference between SAF and SF cases is that, the saturated efficiency in SAF case is much larger than that in SF case. The mechanism is as follows: In SF case, the saturated SOT efficiency is proportional to $1/M_z$, where $M_z$ is the magnetization in z direction of the SF sample under the assist field. However, as Eq. (2) shows, in SAF case, the SOT efficiency is proportional to $1/(M_z^b + M_z^u)$, or $1/(|M_z^b|-|M_z^u|)$ with the same coefficient as in SF case. Because the magnetizations of the upper and the bottom are fully compensated, the small difference $|M_z^b|-|M_z^u|$ is caused by the different tilting angle, which leads to a large saturated efficiency. Even though the incoming spin do not interact with the top layer, still an increased efficiency is seen in our completely compensated SAF system. This is not identical to a ferrimagnetic system where incoming spin acts on both sub-lattices [41,42]. This high saturated efficiency demonstrates that SAF structure is insensitive to magnetic field but is sensitive to current-induced spin torque: magnetic field acts on both upper and bottom layers and manipulate the system only through the Zeeman energy difference between the two layers, but current-induced spin torque acts on the bottom layer mainly.

It is generally considered that an interlayer exchange coupling torque would accelerate SAF domain wall motion [46]. Such analysis is suitable for rapidly moving domain wall. However, in our case where quasi-static depinning analysis is appropriate, the contribution of exchange coupling torque on domain wall driving force vanishes due to the conservation of overall interlayer exchange energy during the rigid domain wall shift. The exchange coupling torque does not drive the domain wall itself, but accelerates the existing motion when other driving force sources like external field and spin torque are already present.

It has also been reported that the exchange coupling in compensated ferrimagnets may drastically increase the SOT effective field [42]. A useful comparison could be made with our SAF case. In the compensated ferrimagnet case, the SOT effective fields first experienced by individual sublattices were considered inversely proportional to the net compensated magnetization $M_S$. Two effective fields on sublattices have opposite directions and constructively rotate the antiparallel magnetizations from macrospin perspective, and such rotation must be equivalent to an even larger uniform effective field, scaling faster than $1/M_S$. The role of exchange coupling here is to keep moments antiparallel and rotating together. In our SAF case, however, the SOT effective field first experienced by the bottom layer is inversely proportional to the bottom layer magnetization, not the net compensated one. The overall effective field experienced by the SAF domain wall, which we measured in experiment, then scales with $1/\Delta M_z$. The role of exchange coupling here is to keep SAF antiparallel domain wall coupled and moving as a whole. Essentially, exchange coupling plays a similar role in SAF as in compensated ferrimagnets, ensuring a larger overall SOT effective field.

Another important feature of the SOT efficiency in our SAF simples is that the saturated efficiency increases with increasing antiferromagnetic coupling. To explain this feature, we need to point out that our model assumes that domain walls are tightly coupled. This is true in strongly antiferromagnetically coupling case, but when the coupling is weaker, domain wall may separate during propagating process, and it would be easier to switch the bottom layer, which is predominant, by magnetic field. However, the difficulty for current induce switching is less dependent on antiferromagnetic coupling strength. Hence, the SOT efficiency would be reduced in weak coupling case, compared to that in strong coupling case.

As Eq. (4) shows, the saturated condition of SOT efficiency in SAF structure is approximately $H_{ext} > H_{DMI}^b + H_{ex}$, since the uniform chirality of the bottom layer domain walls cannot be broken until the external field overcomes the DMI effective field and the exchange coupling field. Such a large in-plane field is indeed a drawback of this work. However, the uniform chirality of the bottom layer domain walls can also be broken by a local exchange pinning field [26], which acts on the bottom layer only. In this case, with a similar method used in Eq. (4), we obtain the required exchange pinning field as $H_{pin} > H_{DMI}^{b} + H_{DMI}^{u}$. Since $H_{DMI}^{u}$ can be much smaller than $H_{ex}$ and $H_{DMI}^{b}$, this required exchange pinning substituting the in-plane field is comparable to the conventional case, which can be exerted by replacing Pt with some antiferromagnetic heavy metal. Therefore, in principle, a field-free switching can be achieved, which would make the SAF structure more competitive.

## 4. CONCLUSIONS

In conclusion, we have demonstrated the spin-orbit torque in perpendicularly magnetized Pt/[Co/Pd]/Ru/[Co/Pd] SAF structure, which shows completely compensated magnetizations and a relatively high exchange coupling field. Although the existence of exchange coupling raises the required assisting external field, the critical current for magnetization switching in the SAF structure is still comparable to the ferromagnetic counterpart because of the high SOT effective field efficiency. The efficient switching of completely compensated SAF might advance magnetic memory devices with high density, high speed, and low power consumption.

**ACKNOWLEDGMENTS**

We acknowledge the support of Beijing Innovation Center for Future Chip (ICFC) and Young Chang Jiang Scholars Program. This work has been supported by the National Key R&D Program of China (Grant Nos. 2017YFB0405704) and the National Natural Science Foundation of China (Grant Nos. 51571128, 51671110 and 51601006).

[1] S. Yuasa, A. Fukushima, T. Nagahama, K. Ando, and Y. Suzuki, Nat. Mater. **3**, 868 (2004).

[2] C. Chappert, A. Fert, and F. N. Van Dau, Nat. Mater. **6**, 813 (2007).

[3] S. Ikeda, K. Miura, H. Yamamoto, K. Mizunuma, H. D. Gan, M. Endo, S. Kanai, J. Hayakawa, F. Matsukura, and H. Ohno, Nat. Mater. **9**, 721 (2010).

[4] S. Parkin, X. Jiang, C. Kaiser, A. Panchula, K. Roche, and M. Samant, Proc. IEEE **91**, 661 (2003).

[5] T. Jungwirth, X. Marti, P. Wadlley, and J. Wunderlich, Nat. Nanotech. **11**, 231 (2016).

[6] C. Song, Y. F. You, X. Z. Chen, X. F. Zhou, Y. Y. Wang and F. Pan, Nanotechnology **29**, 112001 (2018).

[7] B. G. Park, J. Wunderlich, X. Martí, V. Holy, Y. Kurosaki, M. Yamada, H. Yamamoto, A. Nishide, J. Hayakawa, H. Takahashi, A. B. Shick, and T. Jungwirth, Nat. Mater. **10**, 347 (2011).

[8] X. Martí, I. Fina, C. Frontera, J. Liu, P. Wadley, Q. He, R. J. Paull, J. D. Clarkson, J. Kudrnovsky, I. Turek, J. Kuneš, D. Yi, J.-H. Chu, C. T. Nelson, L. You, E. Arenholz, S. Salahuddin, J. Fontcuberta, T. Jungwirth, and R. Ramesh, Nat. Mater. **13**, 367 (2014).

[9] Y. Wang, X. Zhou, C. Song, Y. Yan, S. Zhou, G. Wang, C. Chen, F. Zeng, and F. Pan, Adv. Mater. **27**, 3196 (2015).

[10] Y. Y. Wang, C. Song, G. Y. Wang, J. H. Miao, F. Zeng, and F. Pan, Adv. Funct. Mater. **24**, 6806 (2014).

[11] P. Wadley, B. Howells, J. Železný, C. Andrews, V. Hills, R. P. Campion, V. Novák, K. Olejník, F. Maccherozzi, S. S. Dhesi, S. Y. Martin, T. Wagner, J. Wunderlich, F. Freimuth, Y. Mokrousov, J. Kuneš, J. S. Chauhan, M. J. Grzybowski, A. W. Rushforth, K. W. Edmonds, B. L. Gallagher, and T. Jungwirth, Science **351**, 587 (2016).

[12] S. Y. Bodnar, L. Šmejkal, I. Turek, T. Jungwirth, O. Gomonay, J. Sinova, A. A.

Sapozhnik, H. -J. Elmers, M. Kläui, and M. Jourdan. Nat. Commun. 9, 348 (2018).

[13] M. Meinert, D. Graulich, T. Matalla-Wagner, arXiv:1706.06983 (2017).

[14] X. F. Zhou, J. Zhang, F. Li, X. Z. Chen, G. Y. Shi, Y. Z. Tan, Y. D. Gu, M. S. Saleem, H. Q. Wu, F. Pan, and C. Song, Phys. Rev. Appl. **9**, 054028 (2018).

[15] X. Z. Chen, R. Zarzuela, J. Zhang, C. Song, X. F. Zhou, G. Y. Shi, F. Li, H. A. Zhou, W. J. Jiang, F. Pan, and Y. Tserkovnyak, Phys. Rev. Lett. **120**, 207204 (2018).

[16] S. S. Parkin, Phys. Rev. Lett. **67**, 3598 (1991).

[17] D. Apalkov, B. Dieny, and J. M. Slaughter, Proc. IEEE **104**, 1796 (2016).

[18] J. Hayakawa, S. Ikeda, Y. M. Lee, R. Sasaki, T. Meguro, F. Matsukura, H. Takahashi, and H. Ohno, Jpn. J. Appl. Phys. **45**, L1057 (2006).

[19] C. Yoshida, T. Takenaga, Y. Iba, Y. Yamazaki, H. Noshiro, K. Tsunoda, A. Hatada, M. Nakabayashi, A. Takahashi, M. Aoki, and T. Sugii, IEEE Trans. Magn. **49**, 4363 (2013).

[20] I. M. Miron, K. Garello, G. Gaudin, P.-J. Zermatten, M. V. Costache, S. Auffret, S. Bandiera, B. Rodmacq, A. Schuhl, and P. Gambardella, Nature **476**, 189 (2011).

[21] L. Liu, C. F. Pai, Y. Li, H. W. Tseng, D. C. Ralph, and R. A. Buhram, Science **336**, 555 (2012).

[22] L. Liu, T. Moriyama, D. C. Ralph, and R. A. Buhrman, Phys. Rev. Lett. **106**, 036601 (2011).

[23] P. P. J. Haazen, E. Murè, J. H. Franken, R. Lavrijsen, H. J. M. Swagten, and B. Koopmans, Nat. Mater. **12**, 299 (2013).

[24] S. Emori, U. Bauer, S. M. Ahn, E. Martinez, and G. S. D. Beach, Nat. Mater. **12**, 611 (2013).

[25] G. Yu, P. Upadhyaya, Y. Fan, J. G. Alzate, W. Jiang, K. L. Wong, S. Takei, S. A. Bender, L. T. Chang, Y. Jiang, M. Lang, J. Tang, Y. Wang, Y. Tserkovnyak, P. K. Amiri, and K. L. Wang, Nat. Nanotech. **9**, 548 (2014).

[26] S. Fukami, C. Zhang, S. D. Gupta, A. Kurenkov, and H. Ohno, Nat. Mater. **15**, 535 (2016).

[27] X. P. Qiu, W. Legrand, P. He, Y. Wu, J. W. Yu, R. Ramaswamy, A. Manchon, and H. Yang, Phys. Rev. Lett. **117**, 217206 (2016).

[28] C. Song, B. Cui, F. Li, X. Zhou, and F. Pan, Prog. Mater. Sci. **87**, 33 (2017).

[29] C. Bi, H. Almasi, K. Price, T. Newhouse-Illige, M. Xu, S. R. Allen, X. Fan, and W. Wang, Phys. Rev. B **95**, 104434 (2017).

[30] G. Y. Shi, C. H. Wan, Y. S. Chang, F. Li, X. J. Zhou, P. X. Zhang, J. W. Cai, X. F. Han, F. Pan, and C. Song, Phys. Rev. B **95**, 104435 (2017).

[31] See Supplemental Material at http://link.aps.org/supplemental/ for out-of-plane and in-plane hysteresis loops, SOT switching curves, and 1D micro-magnetism model.

[32] N. Nagaosa, J. Sinova, S. Onoda, A. H. MacDonald, and N. P. Ong, Rev. Mod. Phys. **82**, 1539 (2010).

[33] J. Sinova, S. O. Valenzuela, J. Wunderlich, C. H. Back, and T. Jungwirth, Rev. Mod. Phys. **87**, 1213 (2015).

[34] O. J. Lee, L. Q. Liu, C. F. Pai, H. W. Tseng, Y. Li, D. C. Ralph, and R. A. Buhrman, Phys. Rev. B **89**, 024418 (2014).

[35] G. Yu, P. Upadhyaya, K. L. Wong, W. Jiang, J. G. Alzate, J. Tang, P. K. Amiri, and K. L. Wang, Phys. Rev. B **89**, 104421 (2014).

[36] N. Perez, E. Martinez, L. Torres, S.-H. Woo, S. Emori, and G. S. D. Beach, Appl. Phys. Lett. **104**, 092403 (2014).

[37] J. Torrejon, F. Garcia-Sanchez, T. Taniguchi, J. Sinha, S. Mitani, J. V. Kim, and M. Hayashi, Phys. Rev. B **91**, 214434 (2015).

[38] A. Ghosh, S. Auffret, U. Ebels, and W. E. Bailey, Phys. Rev. Lett. **109**, 127202 (2012).

[39] K. Kondou, H. Sukegawa, S. Mitani, K. Tsukagoshi, and S. Kasai, Appl. Phys. Express

**5**, 073002 (2012).

[40] S. Yakata, Y. Ando, T. Miyazaki, and S. Mizukami, Jpn. J. Appl. Phys. **45**, 3892 (2006).

[41] J. Finley, and L. Liu, Phys. Rev. Appl. **6**, 054001 (2016).

[42] R. Mishra, J. Yu, X. Qiu, M. Motapothula, T. Venkatesan, and H. Yang, Phys. Rev. Lett. **118**, 167201 (2017).

[43] C.-F. Pai, M. Mann, A. J. Tan, and G. S. D. Beach, Phys. Rev. B **93**, 144409 (2016).

[44] J. Han, A. Richardella, S. Siddiqui, J. Finley, N. Samarth, and L. Liu, Phys. Rev. Lett. **119**, 077702 (2017).

[45] S.-G. Je, D.-H. Kim, S.-C. Yoo, B.-C. Min, K.-J. Lee, and S.-B. Choe, Phys. Rev. B **88**, 214401 (2013).

[46] S.-H. Yang, K.-S. Ryu, and S. Parkin, Nat. Nanotech, **10**, 324 (2015).

**Figure Captions**

**FIG. 1.** (a) Schematic of a Ta/Pt/Co/Pd/Co/Ru/Co/Pd/Co/Pd multilayer. (b) Typical optical microscope image of the Hall bar and the measurement configuration.

**FIG. 2.** Normalized out-of-plane hysteresis loops for samples with different Ru thickness: (a) $t_{Ru}$ = 3 Å, (b) $t_{Ru}$ = 6 Å, (c) $t_{Ru}$ = 7 Å, (d) $t_{Ru}$ = 8 Å, and (e) $t_{Ru}$ = 9 Å. (f) Exchange coupling field versus the Ru thickness.

**FIG. 3.** (a) $R_H$ curves measured when sweeping external field along $z$ direction. (b) Current-induced switching with a fixed external field along +y direction. The blue and orange arrows depict the upper and bottom magnetic moments respectively.

**FIG. 4.** Anomalous Hall curves for samples with different Ru thickness: (a) $t_{Ru}$ = 6 Å and (c) $t_{Ru}$ = 9 Å measured when rotating an external field of 3 kOe in yz plane with $I_{dc}$ = ±3 mA. (b) The shift of the angle $\beta$ versus the DC current under $H_{ext}$ = ±3 kOe for $t_{Ru}$ = 6 Å sample. (d) The efficiency of spin-torque as a function of external field for samples with $t_{Ru}$ = 6, 7, 9 Å and a control sample (SF) consists of only the bottom [Co/Pd] layers of our SAF.

**FIG. 5.** Domain structure and domain wall movement direction under zero external field (a) and large external magnetic field (b). Sketch of Stoner-Wohlfarth model (c) and the collective domain wall model (d). (e) Calculated result of SOT efficiency as a function of external field for SAF and SF samples.

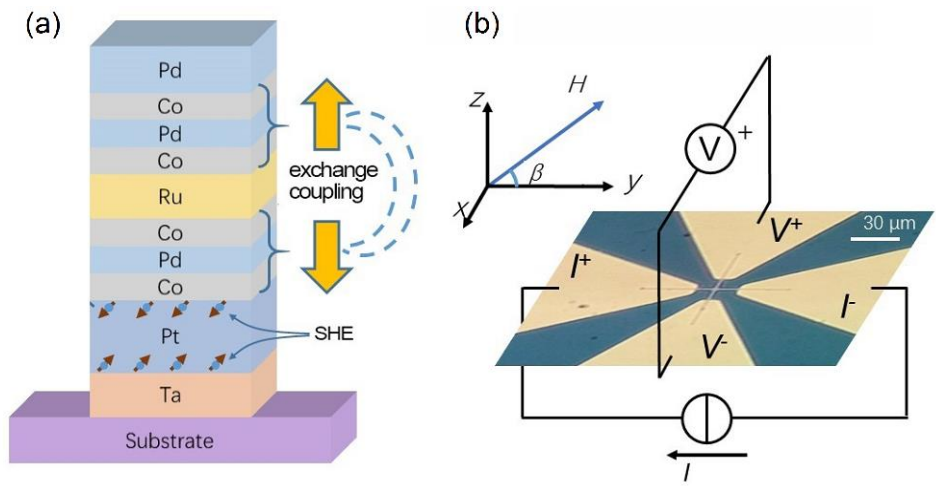

FIG. 1

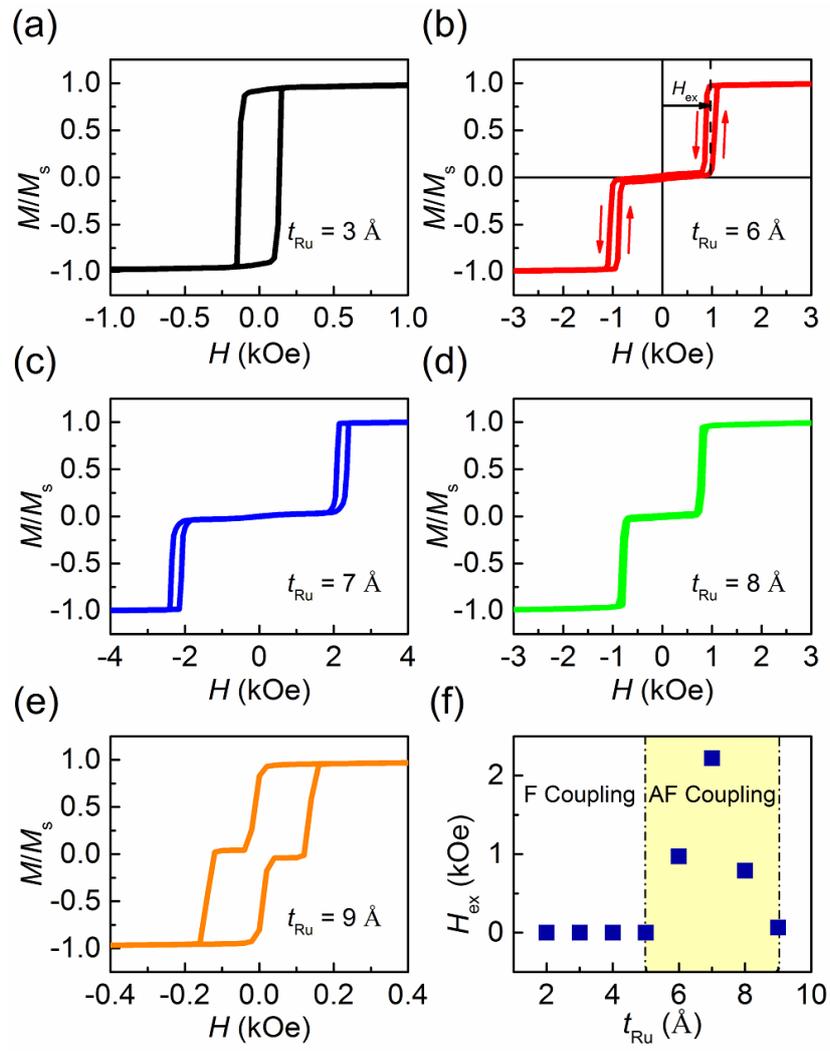

FIG. 2

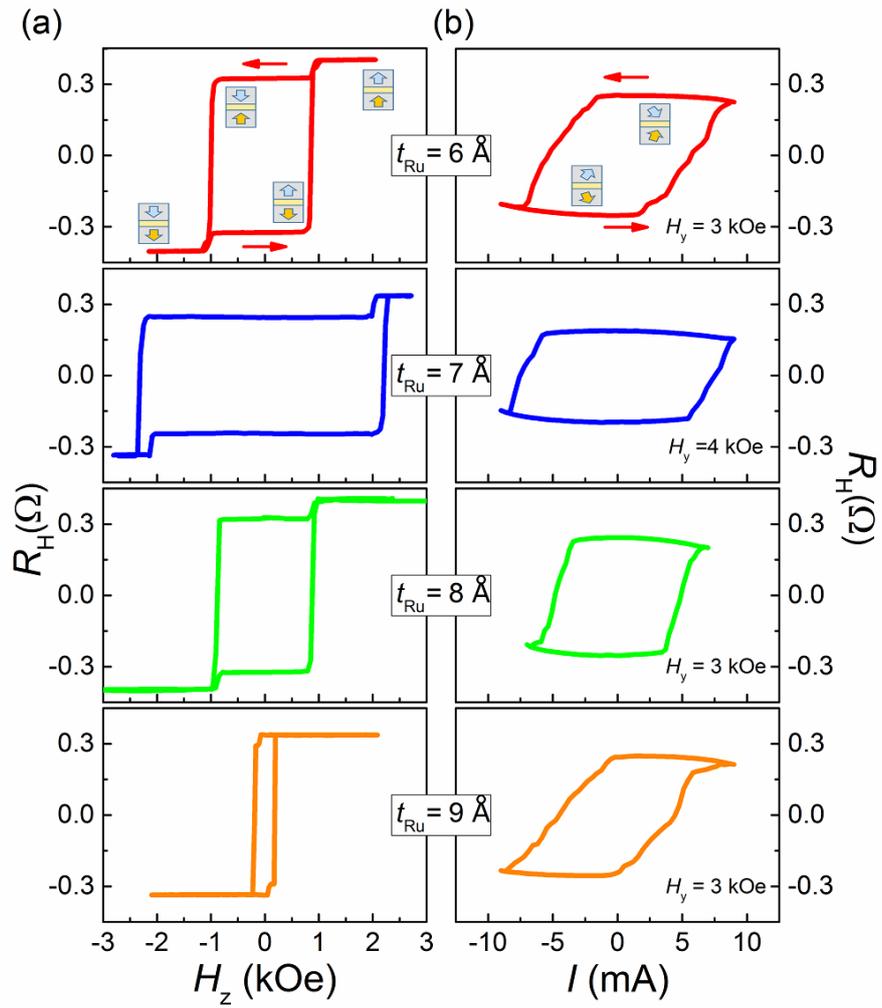

FIG. 3

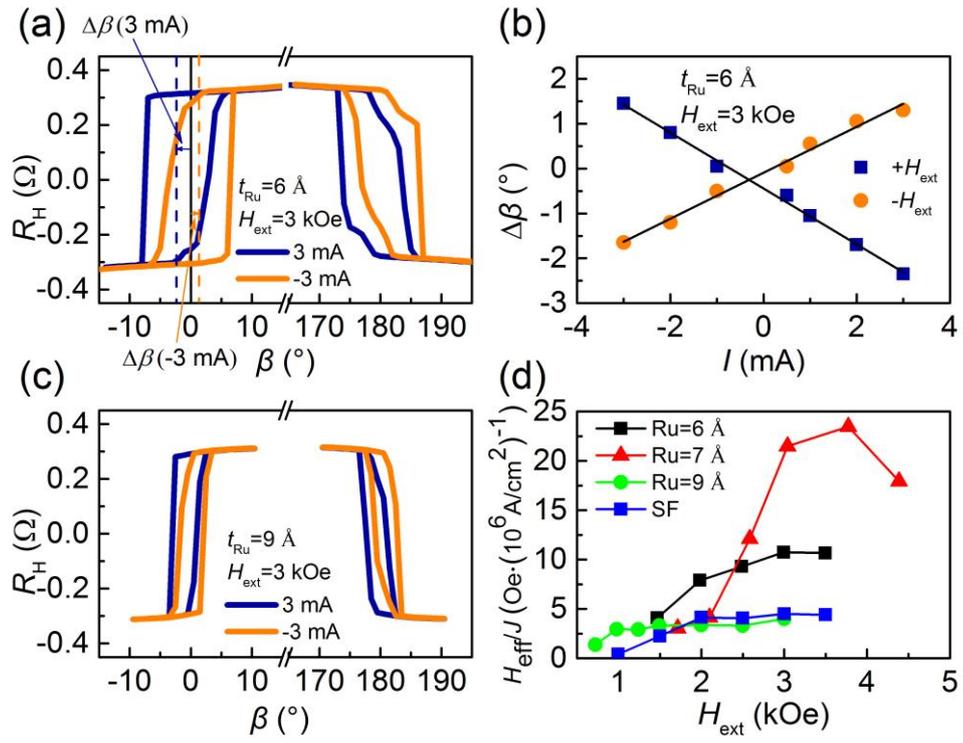

FIG. 4

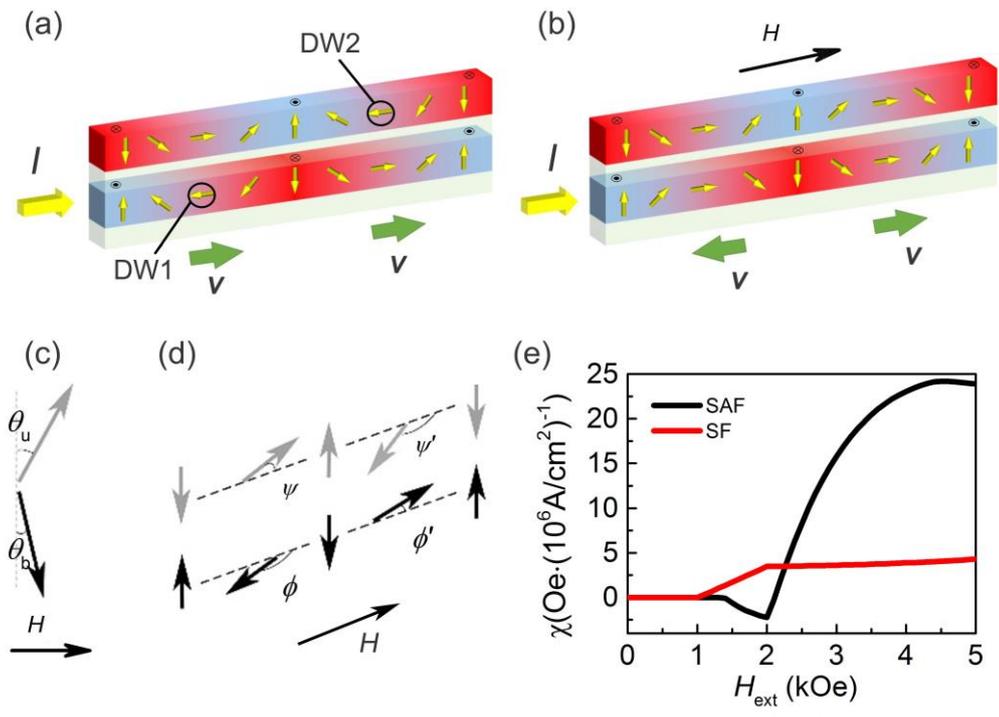

FIG. 5